\documentclass[12pt]{article}   
\usepackage{epsfig,times}  % 17 dec. 2001: pbs avec times
\begin{document}
\thispagestyle{empty} 

%\lhead[\fancyplain{}{\sl }]{\fancyplain{}{\sl }}
%\rhead[\fancyplain{}{\sl }]{\fancyplain{}{\sl }}

%%%%%%%% Pour changer les valeurs par defaut pour taille figure,
%%%%%%%% sinon au-dela d'une hauteur de 134 mm = 70% on est rejete a la fin
 \renewcommand{\topfraction}{.99}      
 \renewcommand{\bottomfraction}{.99} 
 \renewcommand{\textfraction}{.0}

%%%%% Definitions

\newcommand{\nc}{\newcommand}

\nc{\qI}[1]{\section{{#1}}}
\nc{\qA}[1]{\subsection{{#1}}}
\nc{\qun}[1]{\subsubsection{{#1}}}
\nc{\qa}[1]{\paragraph{{#1}}}

            % Enumerations
\def\qbu{\hfill \par \hskip 6mm $ \bullet $ \hskip 2mm}
\def\qee#1{\hfill \par \hskip 6mm #1 \hskip 2 mm}

\nc{\qfoot}[1]{\footnote{{#1}}}
\def\qL{\hfill \break}
\def\qpar{\vskip 2mm plus 0.2mm minus 0.2mm}
\def\qtvi{\vrule height 2pt depth 5pt width 0pt}
\def\qth{\vrule height 12pt depth 0pt width 0pt}
\def\qtb{\vrule height 0pt depth 5pt width 0pt}
\def\tvi{\vrule height 12pt depth 5pt width 0pt}

\def\qparr{ \vskip 1.0mm plus 0.2mm minus 0.2mm \hangindent=10mm
\hangafter=1}

                % Decale UN paragraphe
                % Attention! La double accolade est vitale, sinon tout le
                % est decale (cf TEX p.199)
                % On peut aller a la ligne avec \qL=\hfill \break
                % Par contre ne supporte pas les lignes blanches
\def\qdec#1{\par {\leftskip=2cm {#1} \par}}

   %% Defs specifiques
\def\qdpt{\partial_t}
\def\qdpx{\partial_x}
\def\qddpt{\partial^{2}_{t^2}}
\def\qddpx{\partial^{2}_{x^2}}
\def\qn#1{\eqno \hbox{(#1)}}
\def\qds{\displaystyle}
\def\qw{\widetilde}
\def\qmax{\mathop{\rm Max}}   % Petit livre Tex (p.167)
\def\qmin{\mathop{\rm Min}}   % Petit livre Tex (p.167)

%%%%% End of definitions

\def\qci#1{\parindent=0mm \par \small \parshape=1 1cm 15cm  #1 \par
               \normalsize}

\null
% {\large \it To appear in Physica A}
{\large To appear in the {\it Journal of Economic Interaction and Coordination}}
\vskip 1.5 cm

\centerline{\bf \Large White flight or flight from poverty?}
%\vskip 5mm
%\centerline{\bf \Large }                                      

\vskip 1cm
\centerline{\bf Charles Jego $ ^1 $ }
\vskip 2mm
\centerline{\bf Ecole Polytechnique}
\vskip 10mm

\centerline{\bf Bertrand M. Roehner $ ^2 $ }
\vskip 2mm
\centerline{\bf Institute for Theoretical and High Energy Physics}
\centerline{\bf University Paris 7 }

\vskip 2.5cm

{\bf Abstract}\quad 
The phenomenon of White flight is often illustrated by the case
of Detroit whose population dropped from 1.80 million to 0.95 million
between 1950 and 2000 while at the same time its Black and
Hispanic component grew from 30\% to 85\%. But is this case
really representative? The present paper shows that
the phenomenon of White flight is in fact essentially a flight from poverty. 
As a confirmation, we show that the changes in
White or Black populations are highly correlated which
means that White flight is always paralleled by Black flight
(and Hispanic flight as well).
This broader interpretation of White flight
accounts not only for the case of northern cities such as
Cincinnati, Cleveland or Detroit, but for all population changes at 
county level, provided the population density is higher than a threshold
of about 50 per square-kilometer which corresponds
to moderately urbanized areas (as can be found in states like
Indiana or Virginia for instance).
 
\vskip 8mm
\centerline{November 19, 2005}

%\vskip 8mm
%\centerline{\it Preliminary version, comments are welcome}
\centerline{\it Submitted to the Journal of Economic Interaction and
Coordination}

\vskip 1cm
Key-words: interaction between minorities, White flight, Black flight, poverty,
population migrations.
\vskip 3cm 

1: Charles Jego, Ecole Polytechnique, 91128 Palaiseau Cedex, France.
\qL
\phantom{1: }E-mail: jego@pascal.cpht.polytechnique.fr
\vskip 5mm

2: Bertrand Roehner, LPTHE, University Paris 7, 2 place Jussieu, 
F-75005 Paris, France.
\qL
\phantom{2: }E-mail: roehner@lpthe.jussieu.fr
\qL
\phantom{2: }FAX: 33 1 44 27 79 90

\vfill \eject

\qI{Introduction}

Many ``theories'' have been proposed to ``explain'' White flight.
One can mention for instance the works of Castells (1983), 
Shelling (1971, 1978) or Smith (1987). However, as often occurs in
the social sciences, the theories were proposed almost independently
of actual observations. Two reasons may help to understand why
observation lagged behind. First, in order to be meaningful, the study
of White flight must be carried out at the level of counties or even
the more detailed description level of city-blocks. 
The fact that there are about 5,000 counties in the United States may
explain why before the Internet revolution data at county level
were not easily available. The second difficulty is the fact that one
needs data for different population groups. The very definition of
these groups involves conceptual difficulties%
\qfoot{It must be emphasized that there can
be no scientific definition of a White, Black or Hispanic person. 
Throughout this paper, we rely on the 
statistical definition used  by the U.S. Bureau of the Census which 
is based on a self-identification procedure. For the sake of 
brevity and uniformity, we use systematically the term ``Black''
in preference to other equivalent terms such as ``Afro-American
or ``colored''.}%
.
There are only few countries whose statistical yearbooks
provide data about population components. The United States is probably
the country which publishes the most detailed statistics in this respect.
This is why the study of White flight has largely been confined to this
country. Probably the phenomenon also exists elsewhere as
suggested by the few data which are available for Toronto or London,
but
until more detailed data become available for other countries,
studies of White flight will have to focus on the case of the United States.
\qpar

The study proceeds in five steps.
\qee{1)} First, by taking the specific example 
of the highly urbanized state of New Jersey we explain
how the phenomenon of White flight can be estimated statistically.
\qee{2)} Then, we show that the observations made for New Jersey
can be extended to other states as well provided one focuses on
counties whose population density is above a given threshold.
\qee{3)} We show that White flight is almost as strongly correlated
with poverty as with percentage of minority populations. 
\qee{4)} As poverty is a notion which extends beyond ethnic 
division lines, it is natural to wonder whether or not
different populations groups
have the same behavior with respect to poverty stricken areas.
The evidence shows that the
migration pattern observed for 
white populations is in fact paralleled by similar patterns for
Blacks or Hispanics. White flight, Black flight and Hispanic flight
go hand in hand. Accordingly, White flight appears a rather misleading
denomination; a more 
appropriate one would indeed be flight from poverty.
\qee{5)} In the last section of the paper, we suggest
that this broader interpretation of White flight
can also account for the standard cases of
White flight observed in Cincinnati , Cleveland or Detroit
in the 1950s and 1960s.
\qpar

Before considering the phenomenon at county level it is natural to ask
whether or not it exists at state level. The answer is no.  More
precisely, if one estimates White flight at state level by using the
statistical procedure that we use at county level, one finds a
(non-significant) correlation of -0.15 as compared to correlations
between -0.75 and -0.93 at county level. In other words, at state
level the effect is so weak that it is completely hidden by the
``noise'' due to other shocks and migration factors.

\qI{White flight in New Jersey}

New Jersey is a convenient laboratory for the study of White flight 
because it is a highly urbanized state with a large population. As we
will see later on, population density is a crucial parameter in the study
of white flight. The statistical procedure that we use in order to measure
the intensity of White flight is based on the following steps. 
\qL
For each of the 21 counties of
New Jersey we consider the following variables:
\qee{a)} Total population in 1990: $ P $
\qee{b)} White non-Hispanic (WnH) population in 1990: $ W_1 $
\qee{c)} White non-Hispanic population in 1996: $ W_2 $
\qpar

From these population variables we compute two ratios that we express
as percentages:
$$ \hbox{\qbu Percentage of the population that is not WnH in 1990: } 
x=100(1-W_1/P) $$
$$ \hbox{\qbu Relative percentage annual change in WnH population,
1990-1996: } 
y=100 \left({ 1 \over 6} \right) \frac{W_2-W_1}{W_1} $$

The resulting scatter plot is shown in Fig.~1. 
%%-----------------------------------------------
%%%% Fig.1
  \begin{figure}[tb]
    \centerline{\psfig{width=12cm,figure=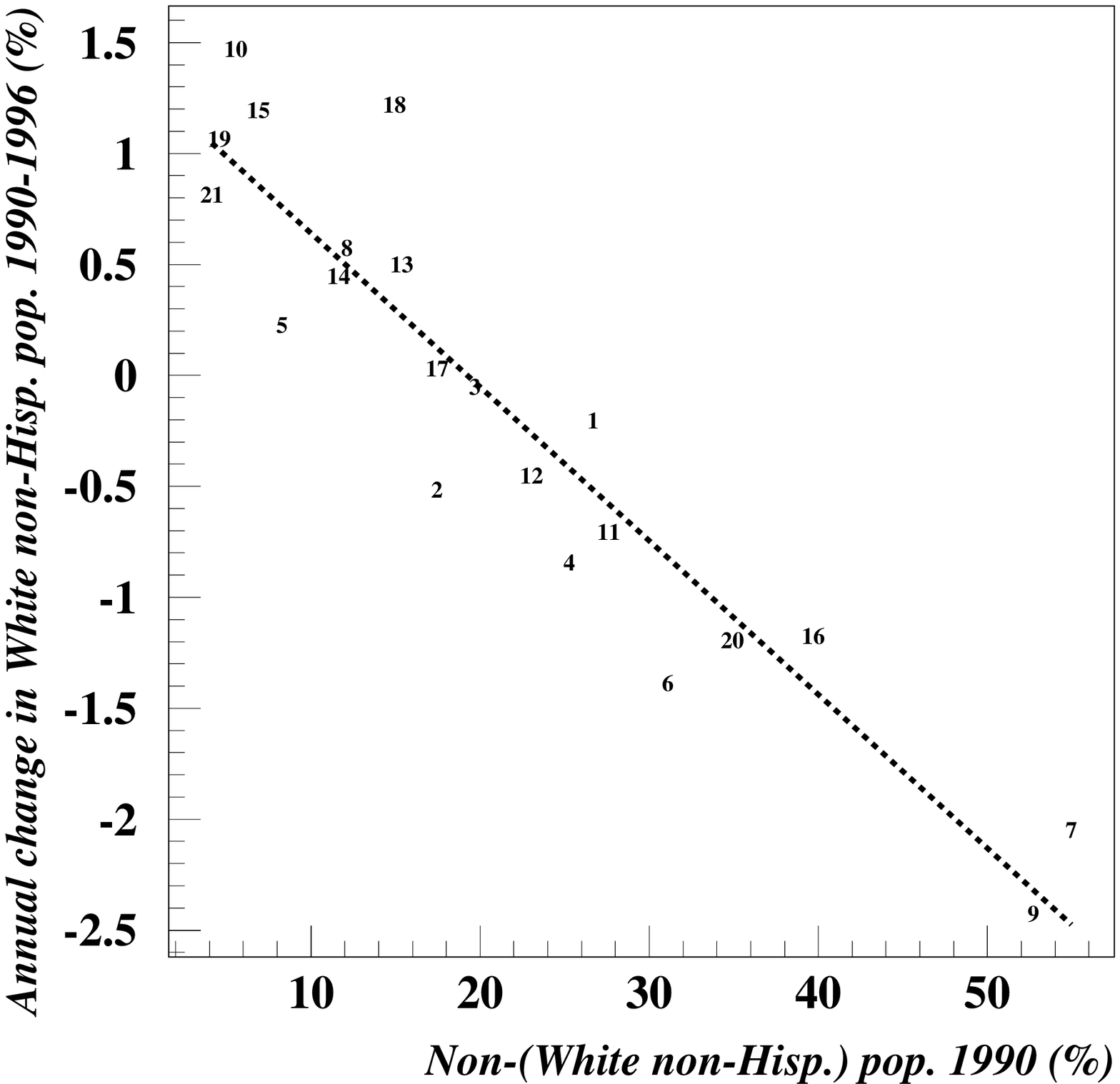}}
    {\bf Fig.~1: White flight in New Jersey.} 
{\small Each number corresponds to one of the 21 counties (listed in
alphabetical order). 
Horizontal axis: percentage of minority populations in 1990;
vertical axis: relative percentage annual change in WnH population. The coefficient
of correlation is -0.93.}
{\small \it Source: USA Counties 1998
(http:// censtats.census.gov/usa/usa.shtml).}
 \end{figure}
%% --------------------------------------------------
There is a strong relationship
between $ x $ and $ y $; the coefficient of correlation is  $ -0.93 $
with a 95\% likelihood confidence interval $ (-0.97, -0.85) $ and the slope
of the regression line is: $ -0.68 \pm 0.11 $.
One may wonder if this result is specific to the time interval
under consideration. The answer is that fairly similar results hold for
previous decades as well: see Table~1.

%%%%%%%%%%%%%%%%%%%%%%%%%%%%%%%
% TABLE 1

\begin{table}[htb]

 \small 

\centerline{\bf Table~1\quad Intensity of white flight in New Jersey
in the time interval 1960-1996}

\vskip 3mm
\hrule
\vskip 0.5mm
\hrule
\vskip 2mm

$$ \matrix{
\tvi 
 \hbox{}  \hfill & 1960-1970 &  1970-1980 & 1980-1990 & 1990-1996 \cr
\noalign{\hrule}
\qth 
 \hbox{Correlation } x-y \hfill & -0.58 & -0.52 & -0.71 & -0.93 \cr
 \hbox{Slope } a \hbox{ of regression line} &  
\qtb
-2.53 \pm 1.6 & -1.58 \pm 1.17 &  0.77 \pm 0.34 & -0.68 \pm 0.11 \cr
\noalign{\hrule}
} $$

\vskip 1.5mm
Notes: \qL
1960-1970, 1970-1980, 1980-1990:
$ x_1=100\% - \hbox{percentage of White population in initial year} $,
$ y_1= $ Relative percentage change in White population over one decade.
\qL
1990-1996:
$ x_2=100\% - \hbox{percentage of White non-Hispanic (WnH)
population} $,
$ y_2= $ Percentage change in WnH population over one decade.
\qL
There are 21 counties in New Jersey; the scatter plot corresponding
to the last time interval is shown in Fig.~1.
Prior to 1990, the data for white
populations also include Hispanics (data for WnH were not available).
This is certainly why
the last correlation is much higher than previous ones; indeed, if one
replaces the ``WnH'' data by 
``White'' data the correlation drops from $ -0.93 $ to $ -0.29 $.
In 1980 the Hispanic population represented 6.1\% as compared
to 11\% in 1996.
The slopes $ a $ of the regression lines in the different time
intervals
were computed on a 10-year basis for all the intervals including the last one
which therefore had to be renormalized by a factor $ 10/6 $.
The fact that $ a $ decreases
in the course of time suggests that the White flight effect
has a tendency to become weaker.
\qL
Source: USA Counties 1998 (http:// censtats.census.gov/usa/usa.shtml).
\vskip 2mm

\hrule
\vskip 0.5mm
\hrule

\normalsize

\end{table}

%% --------------------------------------------------------------

\qI{White flight in other states than New Jersey}

Is White flight specific to New Jersey? That would of course seem 
surprising. As a matter of fact, we find a similar effect in other states
as well. Table 2a summarizes some of the data. The slopes of the regression
lines are comprised between 0.13 and 0.52 (minus sign discarded) and
their average is equal to 0.30. In words, the relationship 
$ |\Delta y| =0.30 |\Delta x| $ means that, for instance, when in a given county
the percentage
of the minorities increases from 20\% to 40\%, then%
\qfoot{In order to give it a more intuitive meaning, the statement is
made in terms of temporal changes, whereas Fig.~1 refers to ensemble
changes at county level. Although the equivalence of temporal 
and ensemble changes cannot be taken for granted, that
assumption is made here as a working hypothesis.}%
:
$$ |\Delta x|=20\% \Longrightarrow |\Delta y| = 6\% $$

which means that
the net change of the WnH population in this county shifts for instance from
8\% to 2\% (i.e. a slowdown in the increase) or from 3\% to -3\% 
(i.e. a shift from an inflow to an outflow).
\qpar

%%%%%%%%%%%%%%%%%%%%%%%%%%%%%%%
% TABLE 2a

\begin{table}[htb]

 \small 

\centerline{\bf Table~2a\quad White flight in various states 1990-1996}

\vskip 3mm
\hrule
\vskip 0.5mm
\hrule
\vskip 2mm

$$ \matrix{
 \hbox{State}  \hfill &   \hbox{Slope of} &   \hbox{Coefficient of} \cr
\tvi 
\hbox{}  \hfill &   \hbox{regression line} & \hbox{correlation} \cr
\noalign{\hrule}
\qth 
\hbox{California}  \hfill & -0.23 \pm 0.09 & -0.55 \cr
\hbox{Georgia}  \hfill & -0.50 \pm 0.07 & -0.71\cr
\hbox{Louisiana}  \hfill & -0.22 \pm 0.08 & -0.53 \cr
\hbox{Maryland}  \hfill & -0.42 \pm 0.17 & -0.71\cr
\hbox{Massachusetts}  \hfill & -0.52 \pm 0.37 & -0.61 \cr
\hbox{New Jersey}  \hfill & -0.41 \pm 0.06 & -0.93 \cr
\hbox{New York}  \hfill & -0.18 \pm 0.04 & -0.70 \cr
\hbox{South Carolina}  \hfill & -0.17 \pm 0.06 & -0.58\cr
\hbox{Texas}  \hfill & -0.17 \pm 0.05 & -0.34\cr
\hbox{Virginia}  \hfill & -0.13 \pm 0.08 & -0.28\cr
\hbox{Washington}  \hfill & -0.32 \pm 0.24 & -0.39 \cr
\hbox{}  \hfill &  & \cr
\qtb
\hbox{\bf Average}  \hfill & \hbox{\bf -0.30} \pm \hbox{\bf 0.04} & \cr
\noalign{\hrule}
} $$

\vskip 1.5mm
Notes: 
$ x=100\% - \hbox{percentage of White non-Hispanic (WnH)
population in 1990} $,
$ y= $ relative percentage change in WnH population, 1990-1996
\qL
In all the cases listed the negative correlation is significant.
The regression slopes refer to the 6 year-long interval 1990-1996;
in order to make them comparable to the slopes for the
decade-long intervals in table 1, they must be multiplied by 10/6.
For instance, the average would become $ 0.30\times (10/6)=0.50 $.
\qL
Source: USA Counties 1998 (http:// censtats.census.gov/usa/usa.shtml).
\vskip 2mm

\hrule
\vskip 0.5mm
\hrule

\normalsize

\end{table}

%% --------------------------------------------------------------

{\bf Apparent exceptions} In some cases there does not seem to
be a clear relationship between $ x $ and $ y $. An example is provided
by Fig.~2a which shows the scatter plot for Ohio. 
%%-----------------------------------------------
%%%% Fig.2a
  \begin{figure}[tb]
    \centerline{\psfig{width=12cm,figure=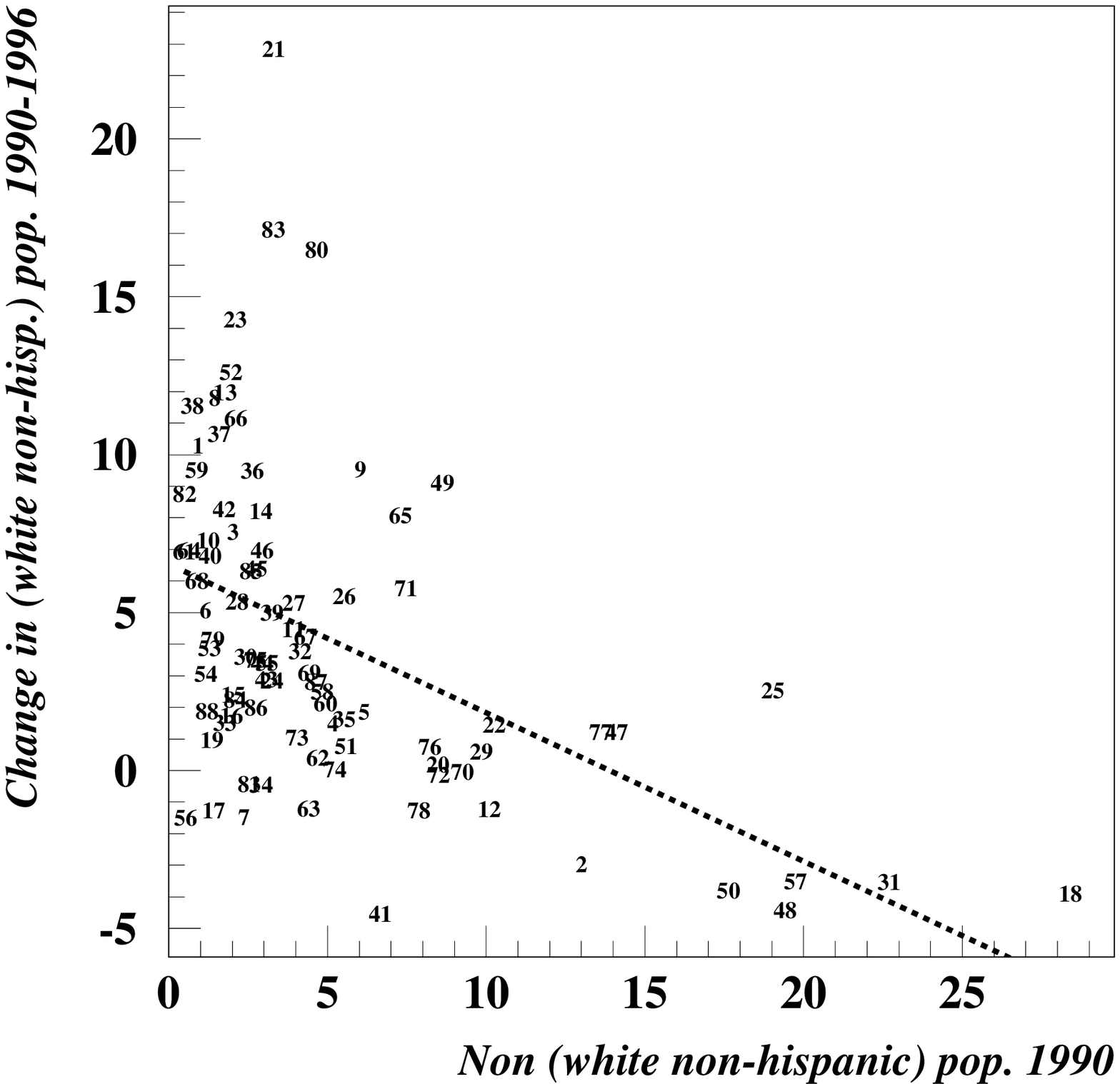}}
    {\bf Fig.~2a: White flight in Ohio.} 
{\small Each number corresponds to one of the 88 counties 
(listed in alphabetical
order). In this graph the population density threshold is 0 which means that
all counties are included. 
The $ x  $ and $ y $ variables are the same as
in Fig.~1. The coefficient of correlation is -0.49
(the 95\% confidence interval is: $(-0.63, -0.31) $ ).}
{\small \it Source: same as in Fig.~1.}
 \end{figure}
%% --------------------------------------------------
%%-----------------------------------------------
%%%% Fig.2b
  \begin{figure}[tb]
    \centerline{\psfig{width=12cm,figure=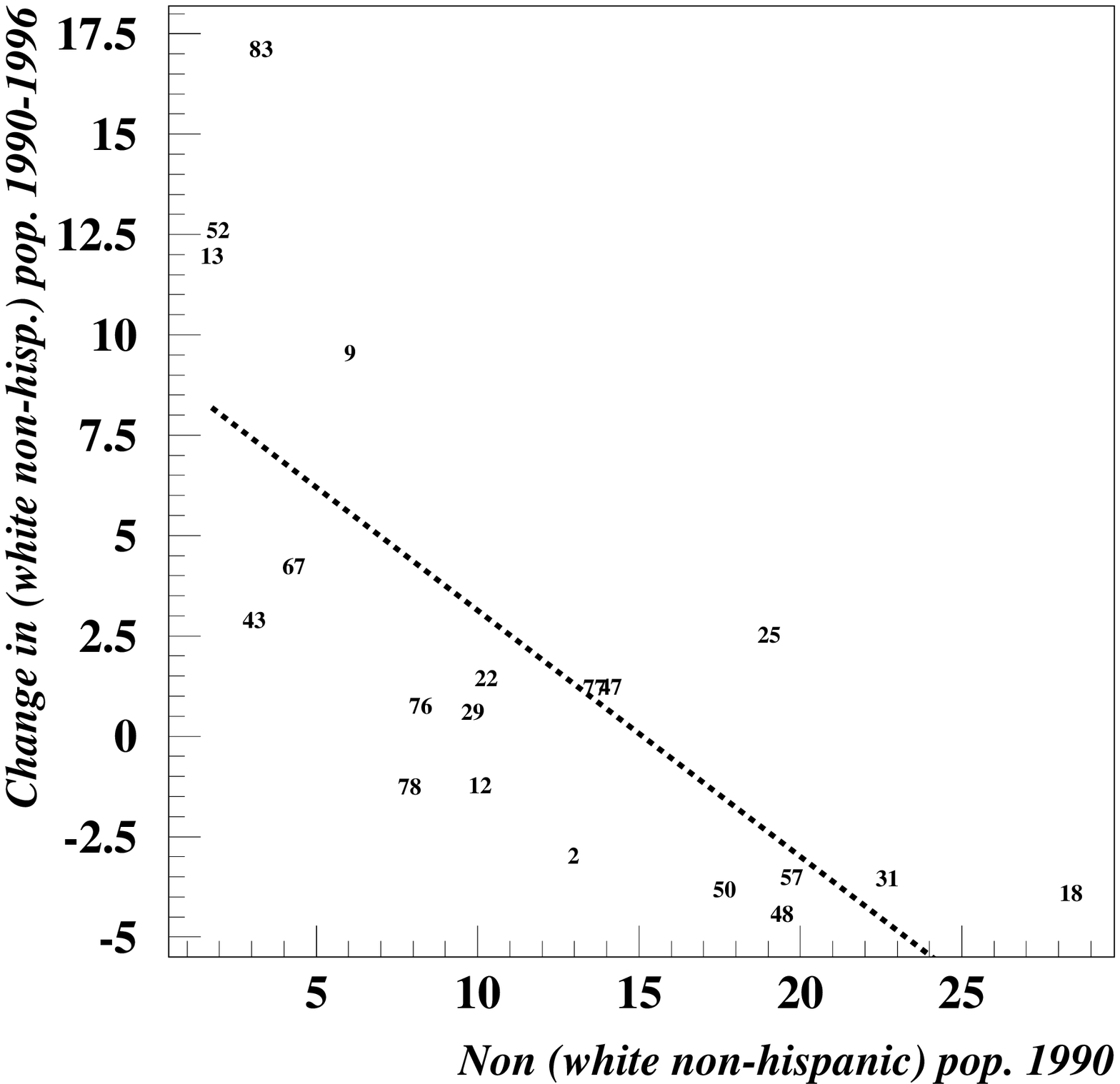}}
    {\bf Fig.~2b: White flight in Ohio.} 
{\small In this graph the population density threshold is 100 people per
square kilometer; there are 20 counties which qualify. 
The $ x  $ and $ y $ variables are the same as in Fig.~1.
The coefficient
of correlation is -0.74 (the 95\% confidence interval is: $ (-0.89, -0.46) $ ).
It is fairly apparent that the relationship between $ x $ and $ y $ is not
linear but rather of the form: $ y=a\ln x + b $; indeed, the coefficient of
correlation of $ \ln x $ and $ y $ is $ -0.83 $ (confidence interval=
$ (-0.93,-0.62) $ ).}
{\small \it Source: same as in Fig.~1.}
 \end{figure}
%% --------------------------------------------------
However, it can be observed that the
correlation increases when the scatter plot is restricted to the counties
whose population density is higher than a given threshold. 
This density effect is documented for several states in Table~2b.

%%%%%%%%%%%%%%%%%%%%%%%%%%%%%%%
% TABLE 2b

\begin{table}[htb]

 \small 

\centerline{\bf Table~2b\quad White flight in areas of low versus
high population density}

\vskip 3mm
\hrule
\vskip 0.5mm
\hrule
\vskip 2mm

$$ \matrix{
\tvi 
 \hbox{State}  \hfill &   \hbox{Number} &   \hbox{Number} &
\hbox{Corr.} & \hbox{Corr.} & \alpha & \alpha \cr
\hbox{}  \hfill &   \hbox{of} &   \hbox{of} &
 &  &  &  \cr
\hbox{}  \hfill &   \hbox{counties} &   \hbox{counties} &
 &  &  &  \cr
\qtb
\hbox{}  \hfill &  D>0 &   D>100 &
 D>0 &  D>100 & D>0  &  D>100 \cr
\noalign{\hrule}
\qth 
\hbox{California}  \hfill &  58 &   12 &
  0.55 &   0.79 &  0.34 &  0.46  \cr
\hbox{Georgia}  \hfill &  159 &  16  &
  0.71 &   0.87 &  0.63 &   0.66 \cr
\hbox{Illinois}  \hfill &  102 &  11 &
  0.20 &  0.59 &  0.001 &  -0.01 \cr
\hbox{Louisiana}  \hfill &  64 &   5 &
  0.53 &  0.84 & 0.33 &  0.13 \cr
\hbox{Maryland}  \hfill &  24 &  9  &
  0.71 &  0.79  & 0.43 &  0.21 \cr
\hbox{Massachusetts}  \hfill &  14 &  10  &
  0.61 &  0.90 & 0.13 &  0.63 \cr
\hbox{New Jersey}  \hfill &  21 &  17  &
  0.93 &  0.92 &  0.85 &  0.80 \cr
\hbox{New York}  \hfill &  62 &  18  &
 0.70  &  0.82 & 0.55 &  0.59 \cr
\hbox{Ohio}  \hfill &  88 &  20  &
 0.49 &  0.74 &  0.31 & 0.46  \cr
\hbox{South Carolina}  \hfill &  46 &  4  &
 0.58 &  0.73 & 0.35 & -0.76  \cr
\hbox{Texas}  \hfill &  254 &  13  &
 0.34 &  0.64 &  0.23 &  0.13 \cr
\hbox{Virginia}  \hfill &  135 &  46  &
 0.28 & 0.30  & 0.11  & 0.01  \cr
\qtb
\hbox{Washington}  \hfill &  39 &  5  &
 0.39  &  0.90 & 0.09 &  0.10 \cr
\hbox{}  \hfill &   &    &
  &   &  &   \cr
\noalign{\hrule}
} $$

\vskip 1.5mm
Notes: For the sake of eliminating negative signs,
all numbers have been replaced by their opposites. 
The two columns labeled
``Corr'' give the coefficient of correlation for the counties with a population
density $ D $ higher than $ 0 $ (i.e. all counties) and 
higher than 100 per square kilometer respectively. Similarly,
the two columns with the heading
$ \alpha $ give the bounds of the confidence intervals (at 95\% likelihood)
which are closest to $ 0 $; roughly speaking, the higher $ \alpha $, the
more significant is the correlation.  
The coefficient of correlation for $ D>0 $ is particularly low when
there are many rural counties (e.g. Louisiana or Texas).
The table shows that an increased
density threshold almost always leads to a higher correlation but that
$ \alpha $ is improved only whenever the state has a substantial number
of high density counties (e.g. Georgia, Massachusetts, New York or
Ohio). Note that, for historical reasons, Virginia's county list contains
41 ``cities'', many of which are in fact small towns, e.g. Bedford City has
a population of 6,073. The non-improvement of $ \alpha $ in this case
clearly shows that what matters is the ``real'' degree of urbanization.
\qL
Source: USA Counties 1998 (http:// censtats.census.gov/usa/usa.shtml).
\vskip 2mm

\hrule
\vskip 0.5mm
\hrule

\normalsize

\end{table}

%% --------------------------------------------------------------

The table shows that the correlation is increased for almost all
states. However, the level of significance is increased only for
states which contain a substantial number of high density counties. A
corollary of this observation is the fact that White flight effect cannot
be observed for states such as Arizona or New Mexico which have almost
only low density counties.

\qI{Influence of poverty}

What are the factors which rule domestic migrations in countries
with an ethnically homogeneous population? One would expect that
economic opportunities are an essential factor. As two extreme 
illustrations one can mention the gold rush toward Alaska after the
discovery of gold in this region or the move of Irish people from
rural counties in Ireland
toward the industrialized areas of Britain during
the crisis of the second half of the nineteenth century%
\qfoot{In the eyes of
people of this time, the Irish were in fact
considered as a separate population group. Their poverty, 
drunkenness, red faces and miserable dwellings gave them an
appearance akin to that of  ``savages'' (in Tocqueville's words (1958)).}%
. 
It is natural to assume that this effect also
exists in the Unites States. To test this hypothesis, we first consider
the case of New Jersey. Fig.~3a is
similar to Fig.~1 except that the 
$ x $-axis variable in Fig.~1 has been replaced by the 
percentage of the population below poverty as defined by the
Census Bureau%
\qfoot{The poverty index is based solely on money income and 
does not reflect non-cash benefits such as food stamps, Medicaid
and public housing. The poverty threshold is updated every
year to reflect changes in the Consumer Price Index (Statistical
Abstract of the United States 2005, p. 424).}%
.
%%-----------------------------------------------
%%%% Fig.3a
  \begin{figure}[tb]
    \centerline{\psfig{width=12cm,figure=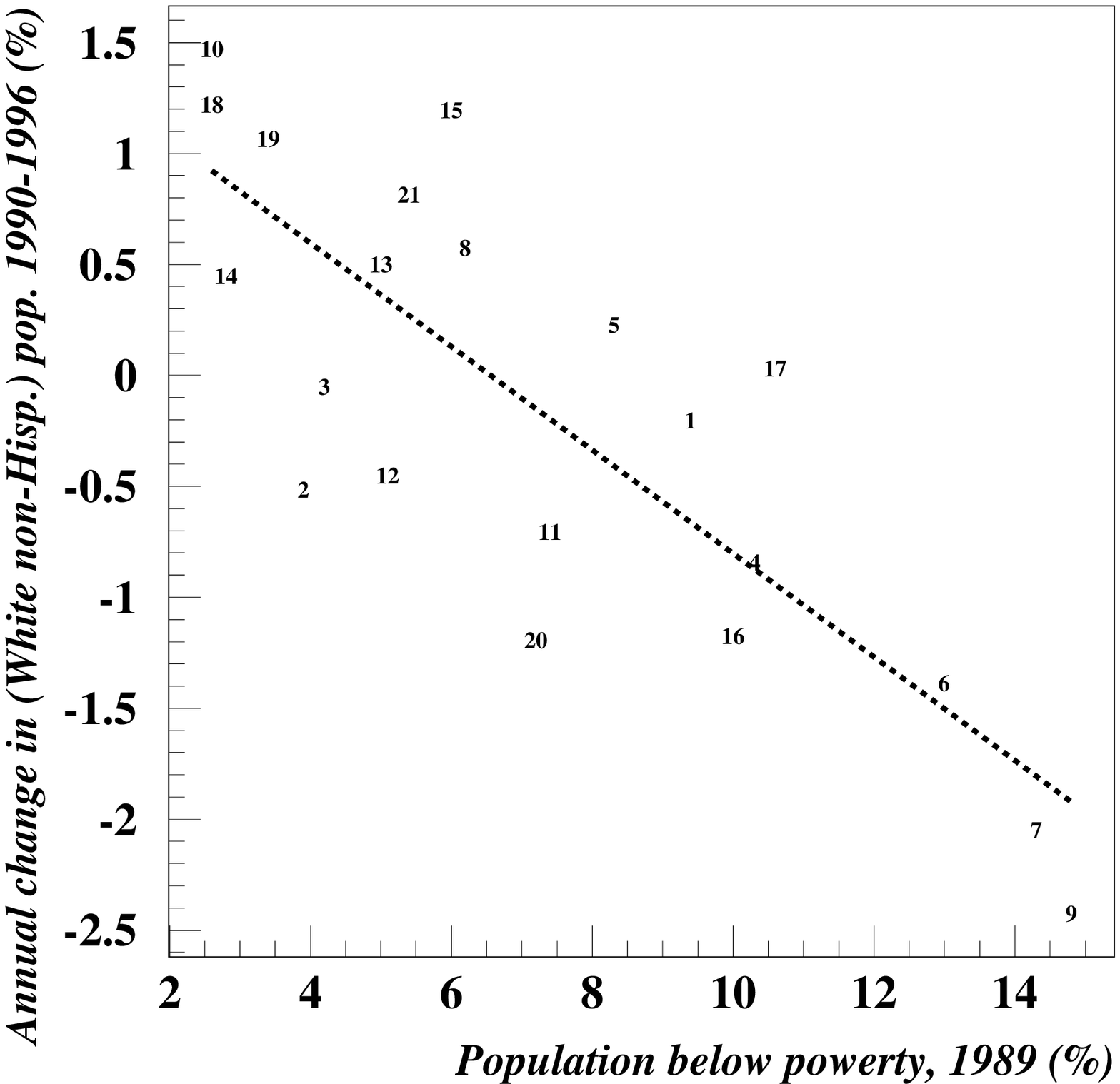}}
    {\bf Fig.~3a: Flight from poverty of White people in New Jersey.} 
{\small Each number corresponds to one of the 21 counties (alphabetical
order). The coefficient of correlation is -0.81. We used the poverty
level in 1989  instead of 1990 because the later was not available on
the County data base.}
{\small \it Source: same as in Fig.~1.}
 \end{figure}
%% --------------------------------------------------
The coefficient of correlation is $ -0.81 $ which confirms that
the poverty effect is almost as strong as the white flight effect
considered in Fig.~1. 
\qpar
{\bf Generalization to other states}\quad Can the results for New Jersey
be extended to other states? Table~3 compares the correlation $ c_1 $
between changes in white population and
minority percentage on the one hand to the correlation $ c_2 $
between the same changes in white population 
and percentage of population below poverty on the other hand.
All the $ c_2 $ correlations are significant with the exception of California
and Louisiana. The reasons behind these exceptions remain an open
question.
On average the $ c_2 $ correlations are only 22\% smaller than the
$ c_1 $ correlations. 
\qpar
%%%%%%%%%%%%%%%%%%%%%%%%%%%%%%%
% TABLE 3

\begin{table}[htb]

 \small 

\centerline{\bf Table~3\quad Is White flight a flight from poverty?}

\vskip 3mm
\hrule
\vskip 0.5mm
\hrule
\vskip 2mm

$$ \matrix{
 \hbox{State}  \hfill &   \hbox{Correlation with} &   \hbox{Correlation with} \cr
\hbox{}  \hfill &   \hbox{minority population} & 
\hbox{population below poverty} \cr
\tvi 
\hbox{}  \hfill &   c_1 & c_2 \cr
\noalign{\hrule}
\qth 
\hbox{California}  \hfill & -0.55&  -0.10\cr
\hbox{Georgia}  \hfill &  -0.71 & -0.66 \cr
\hbox{Louisiana}  \hfill & -0.53 & -0.19  \cr
\hbox{Maryland}  \hfill & -0.71 & -0.67\cr
\hbox{Massachusetts}  \hfill & -0.61 & -0.61 \cr
\hbox{New Jersey}  \hfill & -0.93 & -0.81 \cr
\hbox{New York}  \hfill & -0.70 & -0.51 \cr
\hbox{Washington}  \hfill & -0.39 & -0.24 \cr
\hbox{West Virginia}  \hfill & -0.28 & -0.44 \cr
\hbox{}  \hfill &  & \cr
\qtb
\hbox{\bf Average}  \hfill & \hbox{\bf -0.60}  & \hbox{\bf -0.47} \cr
\noalign{\hrule}
} $$

\vskip 1.5mm
Notes: The $ c_1 $ column gives the correlation between the
same variables as in table 2. The $ c_2 $ column gives the
correlation between $ x= $ Percentage of population below  poverty
 and the same $ y $ variable as in table 2. All the $ c_2 $ correlation
coefficients are significant except those of California and
Louisiana.
\qL
Source: USA Counties 1998 (http:// censtats.census.gov/usa/usa.shtml).
\vskip 2mm

\hrule
\vskip 0.5mm
\hrule

\normalsize

\end{table}

%% --------------------------------------------------------------

\qpar

How are these two effects related? 
One can note that in the 
United States the income of Whites is higher than
the income of Black or Hispanic minorities. In 1990, the
median income of Whites was 1.7 times higher than the median income
of Blacks and 1.4 times higher than the median income of  Hispanics. Therefore
one is not really surprised that the two effects seem to overlap.
Naturally, one would like to understand better the relationship between
these two effects and whether one is a consequence of the other.
One way to explore this issue 
is to study the migrations of Black people.
For the case of New Jersey this is done in Fig.~3b. 
%%-----------------------------------------------
%%%% Fig.3b
  \begin{figure}[tb]
    \centerline{\psfig{width=12cm,figure=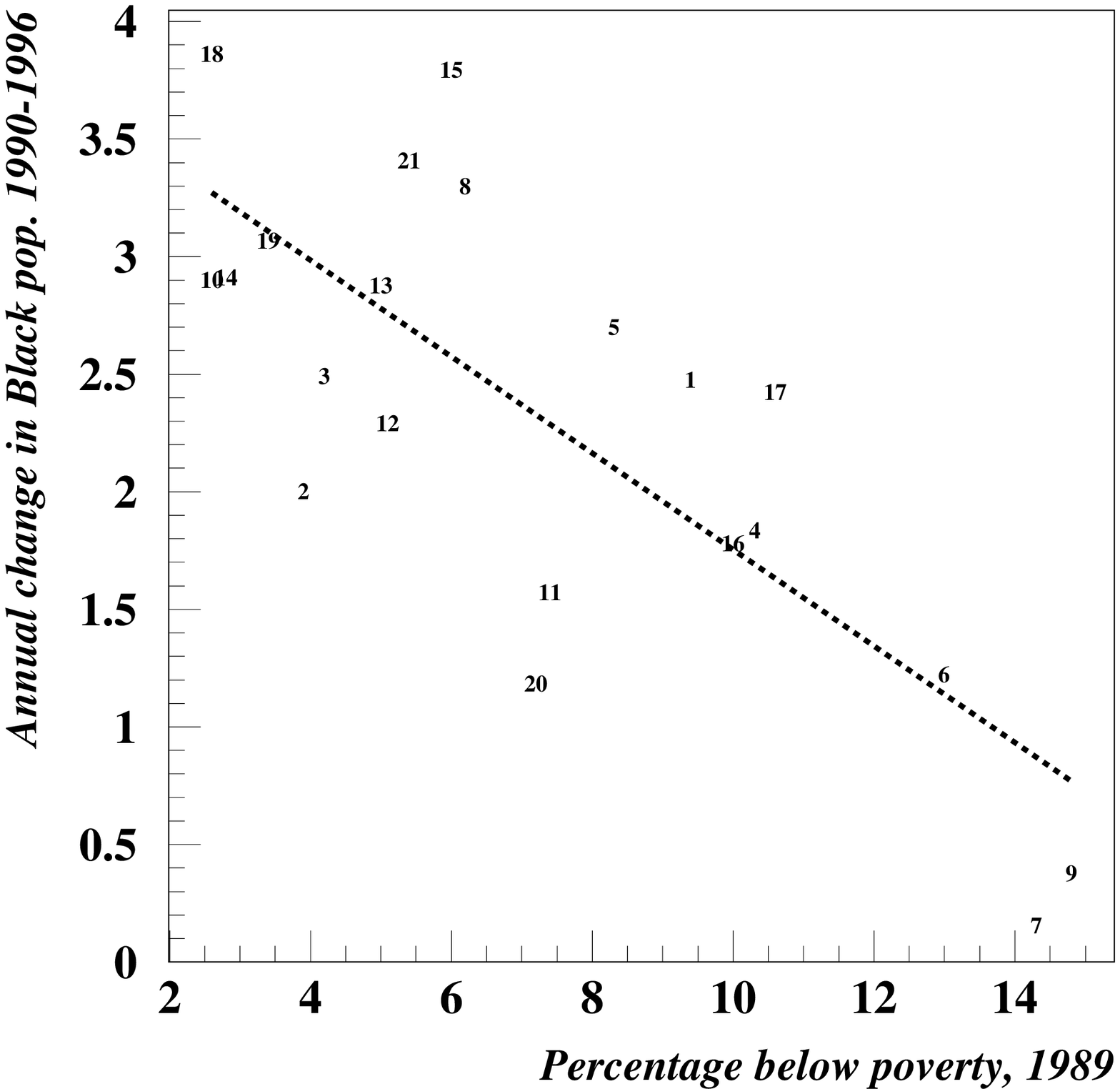}}
    {\bf Fig.~3b: Flight from poverty of Black people in New Jersey.} 
{\small Each number corresponds to one of the 21 counties 
(listed in alphabetical
order). The coefficient of correlation is -0.76.}
{\small \it Source: same as in Fig.~1.}
 \end{figure}
%% --------------------------------------------------
We see that there is a significant Black flight. The correlation is
$ -0.76 $ as compared to $ -0.93 $ in Fig.~1.
As a matter of fact the two graphs are very similar.
It is of interest to observe more closely
the counties at the two opposite ends 
of the horizontal axis.
\qbu The counties numbered 7 and 9 (i.e. Essex and Hudson respectively)
have a high minority percentage but also a high poverty percentage. 
As a result, Whites have been leaving these counties at an annual rate
of about 2.20\%, 
whereas the Black population increased at an annual
rate of only 0.28\% as compared to 1.40\% for the whole state of New Jersey.
In short, there has been a White flight as well as a Black flight 
away from Essex and Hudson counties%
\qfoot{In a general way, in the present study, we do not try 
to distinguish between the natural increase (due
to the surplus of births over deaths) and migration balance.
If we try to do it here, 
broadly speaking, this can be done on the basis that in New Jersey
the annual natural increase 1990-1996 was $ 0.62\% $ for the White
population and $ 1.00\% $ for the Black population.
This means that
for the White population migrations represented 
$ -2.20\%-0.62\%=-2.82\% $, whereas for the
Black population they represented $ 0.28\%-1.00\%=-0.72\% $.
The figures show that there was indeed a Black flight which paralleled
the White flight. }%
. 
\qbu At the other end of the income spectrum, the counties numbered
18 and 19 (i.e. Somerset and Sussex respectively) have a minority percentage
of 10\% and 
have less than 3\% of their population under poverty level. Their White
component has been increasing at an annual rate of
about one percent  and their Black component
at an average rate of 3.5 \%.
\qpar

This suggests that the
so-called White flight is not specific to Whites but is shared by
other components of the population as well and that it
has more to do with economic opportunities
than with interactions between different communities. 
Further insight can be gained by discussing
in more detail the case of West Virginia
(mentioned in Table~3) which is of special interest.
This state has few 
minority people: 3.1\% Blacks, 0.6\% Hispanics and 0.5\% Asians.
Therefore, one would not expect that the minorities are the force
which drives the movements of the White population. This is indeed
confirmed by the results in Table~3: the $ c_1 $ correlation is only
$ -0.28 $ with a 95\% confidence interval $ (-0.51,-0.02) $ which shows
that it is barely significant. On the other hand, with an average below
poverty percentage of 20\% (1990), West Virginia is one of the poorest
American states. As a result, one is not surprised that it is poverty
which is the driving force of the movements of the WnH population as
confirmed by a $ c_2 $  correlation of $ -0.44 $ (i.e. 57\% higher than the 
previous $ c_1 $ correlation of $ -0.28 $).
\qpar

The fact that White flight is paralleled by Black and Hispanic
flights is confirmed by examining the correlations between the 
population changes of these three population components.
This is the purpose of the next section.

\qI{White flight, Black flight, Hispanic flight}

Once again, we begin by examining the case of New Jersey. Fig.~3c shows
that the changes in Black and White non-Hispanic population changes
are highly correlated.
%%-----------------------------------------------
%%%% Fig.3c
  \begin{figure}[tb]
    \centerline{\psfig{width=12cm,figure=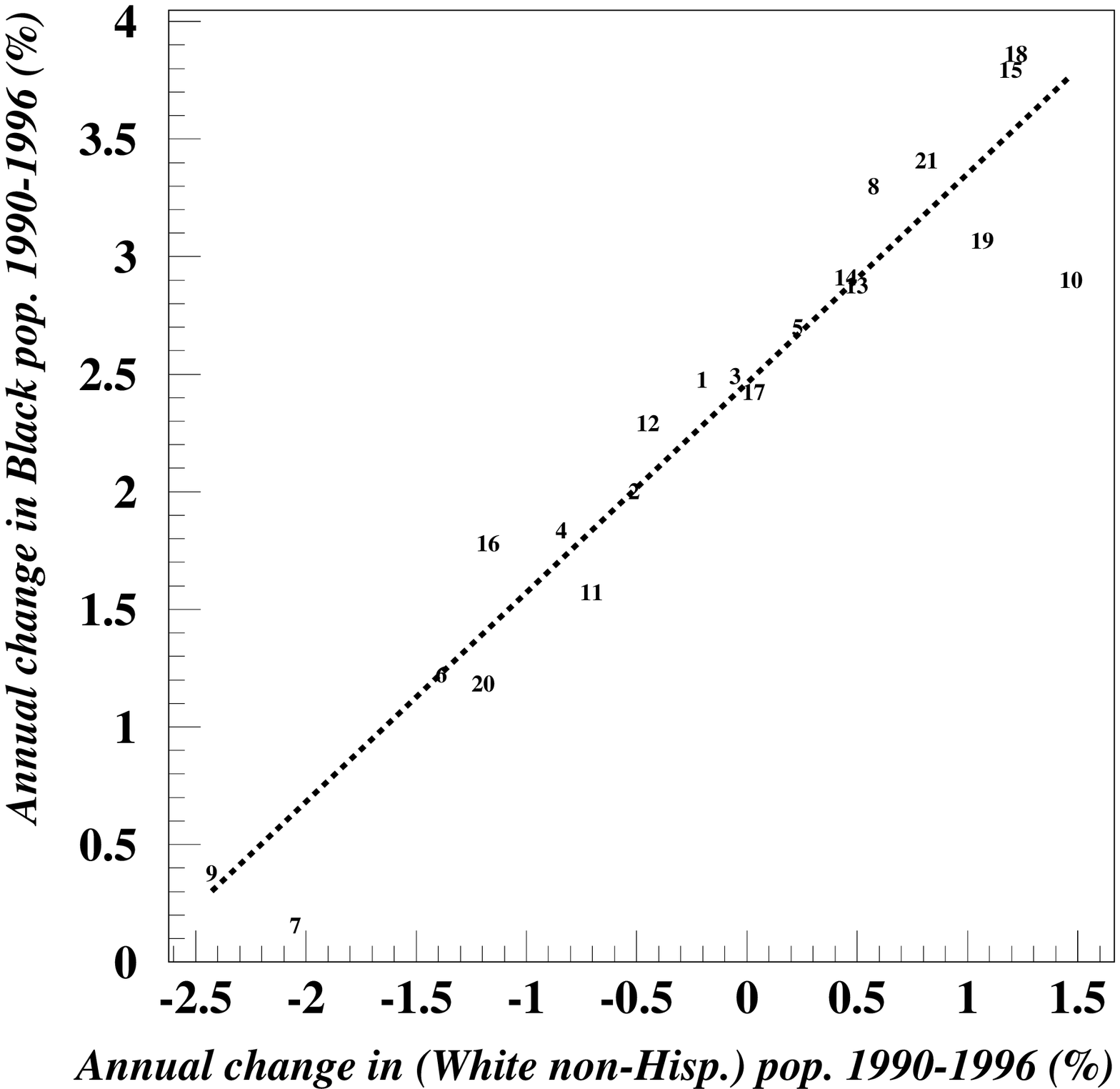}}
    {\bf Fig.~3c: Relationship between white non-hispanic flight and
black flight in New Jersey.} 
{\small Each number corresponds to one of the 21 counties (alphabetical
order). The coefficient of correlation is $ 0.95 $ and the slope of the
regression line is $ 0.89 \pm 0.12 $.}
{\small \it Source: same as in Fig.~1.}
 \end{figure}
%% --------------------------------------------------
The coefficient of correlation $ c $ is equal to $ 0.95 $ and the slope
of the regression line is: $ a =0.89 \pm 0.12 $, which means that
$ \Delta y \sim \Delta x $. For the Hispanic and White non-Hispanic
population changes the correlation is even higher with $ c=0.99 $ and
$ a=1.25 \pm 0.07 $. The same findings hold for other states as well 
at least above a population density threshold of about 50 per 
square kilometer (see Table~4).

%%%%%%%%%%%%%%%%%%%%%%%%%%%%%%%
% TABLE 4

\begin{table}[htb]

 \small 

\centerline{\bf Table~4\quad White flight, Black flight and Hispanic
  flight, 1990-1996}

\vskip 3mm
\hrule
\vskip 0.5mm
\hrule
\vskip 2mm

$$ \matrix{
\tvi 
 \hbox{State}  \hfill &   \hbox{WnH-B} &   \hbox{WnH-H} \cr
\noalign{\hrule}
\qth 
\hbox{California}  \hfill &  &  \cr
\quad \hbox{All counties}  \hfill & 0.67 &  0.90 \cr
\quad \hbox{Counties for which } D>50  \hfill & 0.85 &  0.96 \cr
\hbox{Maryland}  \hfill &  &  \cr
\quad \hbox{All counties}    \hfill & 0.90 &  0.90 \cr
\quad \hbox{Counties for which } D>50  \hfill & 0.95 &  0.98 \cr
\hbox{Massachusetts}  \hfill &  &  \cr
\quad \hbox{All counties}    \hfill & 0.95 &  0.96 \cr
\quad \hbox{Counties for which } D>50  \hfill & 0.93 &  0.97 \cr
\hbox{New Jersey}  \hfill &  &  \cr
\quad  \hbox{All counties}   \hfill & 0.95 &  0.99 \cr
\quad \hbox{Counties for which } D>50  \hfill & 0.95 &  0.99 \cr
\hbox{Texas}  \hfill &  &  \cr
\quad  \hbox{All counties}   \hfill & 0.09 &  0.43 \cr
\quad \hbox{Counties for which } D>50  \hfill & 0.85 &  0.98 \cr
\quad \hbox{} D>50  \hfill &  &   \cr
\hbox{\bf Average, all counties }  \hfill & \hbox{\bf 0.71} &  \hbox{\bf 0.84}  \cr
\qtb
\hbox{\bf Average, counties } D>50 \hfill & \hbox{\bf 0.91} &  \hbox{\bf 0.98} \cr
\noalign{\hrule}
} $$

\vskip 1.5mm
Notes: WnH means White non-Hispanic, B means Black,
H means Hispanic. The WnH-B column gives the correlations
between $ x= $ change in WnH population, $ y_1=$ change in
B population. The WnH-H column gives the correlations
between the same $ x $ variable and $ y_2= $ Change in
H population. $ D $ denotes the population density expressed
in population per square kilometer. The low correlation in Texas 
which is obtained when
all counties are included are due to the fact that Texas has 
a great number of rural counties with small populations
and densities. Only
11\% of the 254 counties have a population density over 50 per
square kilometer. Naturally, counties with small populations can
experience very large population changes. Thus, for instance,
the county of Lasalle (total 1990 population of 5,254
and density=1.4 per sq. km) experienced
a 845\% increase in its black population which grew from 53 in 1990
to 501 in 1996. Such large changes result in a huge dispersion and
thus in a low correlation. 
When these rural counties are left aside, the correlation
becomes much higher. 
\qL
Source: USA Counties 1998 (http:// censtats.census.gov/usa/usa.shtml).
\vskip 2mm

\hrule
\vskip 0.5mm
\hrule

\normalsize

\end{table}

%% --------------------------------------------------------------

\qI{Conclusion and future prospects}

At first sight the previous observation seems to be at variance with
the standard mechanism of White flight. For cities such as Cleveland
or Detroit, it is usually described as a three stage process.
(i) Move of Blacks to northern industrial cities during
World War II and concurrent move of White people to 
suburban areas%
\qfoot{This move was not entirely ``spontaneous'' but was 
encouraged by the availability of affordable mortgages for new
homes in the suburbs and by the development of highways.
It can be noted that both mortgages and highway constructions
were subsidized by the federal government.}%
. 
These opposite flows of population resulted in the
replacement of high or medium wage earners by low wage earners.
(ii) Fall in municipal revenue and accompanying decline in
public services. 
(iii) Decline in overall city population. 
\qpar

This description is not consistent with the findings of this paper.
Yet, as the evidence presented in this paper
pertains to the period 1990-1996, it can be argued that it cannot shed 
new light on what happened in the 1950s. This is certainly true
although it is tempting, at least as a working hypothesis,
to posit that the movements of population are ruled by
mechanisms which are fairly robust in the course of time. 
\qpar

Apart from Cleveland and Detroit, does the above mechanism 
apply to other cities?
By considering other large cities, one comes to the conclusion that
this mechanism does not have a broad validity.
For instance, a growing minority population does not necessarily
result in an overall population decline. Although this was indeed the
case in Cincinnati, Cleveland or Detroit, it is by no means a general
rule as shown by the following counter-examples:
\qbu In New Orleans the Black population increased approximately
in the same way as in Detroit, i.e. from about 33\% in 1950 to about
60\% in 1980. Yet, in contrast to Detroit where the total population
fell by 35\%, it remained almost unchanged in New Orleans.
\qbu In Los Angeles, the share of the minority population increased
from 16\% in 1960 to about 60\% in 2000; yet the total population increased
by 48\% instead of decreasing.
\qbu In Boston, the minorities represented only 28\% in 1980. Yet, from
1950 to 1980 the total population decreased by 30\%, almost as rapidly
as in Detroit. A similar but less known example is provided by the city
of Charleston in West Virginia. With a population of only
53,000 in 2000 it is a small city yet the largest
in the state. Its minority population represents 16\% (15\% Black and
1\% Hispanic). Yet, between 1950 and 2000 its population dropped by 27\%. 
\qpar

According to the observations made in this paper, {\it all} population
components tend to converge toward  places which 
offer adequate employment and to avoid
places with little opportunities. As a matter of fact, this mechanism
can also explain typical white flight cases such as
Cleveland or Detroit. In the 1950s these cities suffered from several
handicaps. 
\qbu In the U.S. as well as in European countries the second half
of the 20th century was marked by a migration trend toward sunny
areas, for example toward the states of
Arizona, California, Florida or Texas. As a consequence,
the Midwest states lost their attractiveness unless they had specific
assets.
\qbu In a time of declining
industrial activity, the Cleveland-Cincinnati-Detroit area
had the misfortune that its economy was mainly based on
shrinking industrial sectors. 
\qbu Moreover, these cities had
no renowned universities and their centers had none of the assets of
cities such as Boston, Manhattan or San Francisco. 
\qpar
In short, the decline of this region
was probably inevitable even if the population had been
mono-ethnic. As a matter of fact,
similar declines were experienced in
regions which had the same handicaps, for instance the
Birmingham-Liverpool area in Britain or the regions devoted
to heavy industry in the north and north-east of France.
\qpar

In conclusion, the following picture emerges from this study.
The different population components have basically the same 
behavior
in the sense that they are attracted by or kept away from the same areas.
However, due to their higher income, White people have
a greater {\it mobility}. They 
can afford a home in suburban areas as well as adequate means of
transportation whereas Black people remain trapped in areas
in which they arrived in a time of good opportunities but which they
cannot leave as quickly as whites once these opportunities have
vanished.
\qpar

\count101=0  \ifnum\count101=1

There is a question that we did not address so far:
how do the different components of the population
compare in their ability to create
new wealth and how is this ability affected by the contacts between
different cultures and ways of life?
Although a question of fundamental importance,
it can hardly be answered in a short-term perspective. It seems that
in periods of time marked by a globalization process,
the dominant opinion is to assume
that the linkage between two cultures is necessarily profitable
to both of them. Yet, this belief is hardly confirmed by
historical evidence.
The point can be illustrated by the globalization process which occurred
after the discovery of the New World by Christopher Columbus.
\qpar

When Juan de Ampues, Nikolaus Federmann (in Venezuela)%
\qfoot{The territory of New Grenada which corresponds approximately
to present-day Venezuela, was granted by Emperor Charles V to
the Welsers, the great Augsburg banking firm to which he was
heavily indebted. Federmann was an agent of the Welsers.
The government of the Welsers was marked by ruthless exploitation
of the Indians (Langer 1968, p. 529).}%
,
Francisco Pizarro (in Peru and Bolivia) and Hernando Cortes (in Mexico)
came into contact with the peoples of Latin America, they found
vibrant civilizations
characterized by impressive architectural achievements and whose
craftsmanship made the
admiration of Spanish courtiers when they discovered  the magnificent
items sent back by the explorers. 
Unfortunately, the Inca and Aztec civilizations did not thrive
in the new environment. The parallel with other globalization episodes
such as those of the late nineteenth or late twentieth centuries 
becomes more convincing when one realizes that 
the main objective of Federmann, Pizarro
or Cortes was not military conquest but 
``to extract rents from the new territories'' as clearly
stated in a letter of Charles V to Cortes dated  June 26, 1523
(Duverger, 2001). 
\qpar

The purpose of this historical example is to emphasize 
that the interchange between different cultures and ways of life
is {\it not} necessarily mutually beneficial. We do not yet know the
rules which govern the process of cross-fertilization nor do we
know the time constants of the adaptation process.
In their general
form, these are difficult multi-faceted issues, 
yet, at the level of broad
orders of magnitude,  there are perhaps some
simple rules which would at least enable us to better understand
why the incorporation of Latin America into the world economy
in the 16th century turned out to be such a disaster. 
What makes the problem amenable to comparative tests is the fact that
apart from Latin America there are many other
examples; one can mention for instance the incorporation
of:
\qbu Bulgaria and Romania into the Ottoman empire, 
\qbu India into the British Empire, 
\qbu Madagascar into the French Empire.

\fi

\qpar
{\bf Acknowledgments}\quad A preliminary version of this paper
was presented at the ``Workshop of Geographers and
Econophysicists (Paris, May 26$^{th}$ 2004)''; many thanks to all participants
for their comments and especially to Renaud Le Goix, Denise Pumain and 
G\'erard Weisbuch.

%\vfill \eject

\vskip 5mm

{\bf \large References}

\qparr
Castells (M.) 1983: Cultural identity, sexual liberation and urban
structure: the gay community in San Francisco. in: M. Castells, the
city and the grassroots: a cross-cultural theory of urban and social
movements. 
Edward Arnold, London.

\qparr
Duverger (C.) 2001: Cortes. Fayard, Paris.

\qparr
Langer (W.L.) ed. 1968: An encyclopedia of world history.
Houghton Mifflin Company, Boston.

\qparr
Schelling (T.S.) 1971: Dynamic models of segregation.
Journal of Mathematical Sociology 1, 143-186.

\qparr
Schelling (T.S.) 1978: Micromotives and macrobehavior.
W.W. Norton, New York.

\qparr
Smith (M.) 1987: Gentrification and the rent-gap.
Annals of the Association of American Geographers 77, 3, 462-465.

\qparr
Tocqueville (A. de) 1958: Journeys to England and Ireland. Faber and
Faber, London. 

\end{document}